\documentclass[usenatbib]{mnras}
\usepackage[utf8]{inputenc}
\usepackage[english]{babel}
\usepackage[T1]{fontenc}
\usepackage{hyperref}
\usepackage{ae,aecompl}
\usepackage{amsmath}
\usepackage{dsfont}
\usepackage{amssymb}
\usepackage{gensymb}
\usepackage{graphicx}
\usepackage{natbib}
\usepackage{url}
\usepackage{tikz}
\usetikzlibrary{arrows,shapes}
\usepackage{rotating}
\usepackage{longtable}
\usepackage{multirow}
\usepackage{ulem}
\usepackage{listings}
\usepackage{commath}
\usepackage{booktabs}

\newcommand{\ME}{M$_\oplus$}

\title{Could Uranus and Neptune form by collisions of planetary embryos?}
\author[Chau et al.]
{Alice Chau$^1$\thanks{achau@physik.uzh.ch},
 Christian Reinhardt$^1$,								 
 Andr\'e Izidoro$^2$,
 Joachim Stadel$^1$,
 Ravit Helled$^1$
 \\
  $^{1}$Institute for Computational Science, University of Zurich, Winterthurerstrasse 190, 8057 Zurich, Switzerland
  \\
  $^2$ Department of Earth, Environmental and Planetary Sciences, MS 126, Rice University, Houston, TX 77005, USA}

\begin{document}
\label{firstpage}
\pagerange{\pageref{firstpage}--\pageref{lastpage}}
\maketitle

\begin{abstract}
The origin of Uranus and Neptune remains a challenge for planet formation models. A potential explanation is that the planets formed from a population of a few planetary embryos with masses of a few Earth masses which formed beyond Saturn's orbit and migrated inwards. These embryos can collide and merge to form Uranus and Neptune.
In this work we revisit this formation scenario and study the outcomes of such collisions using 3D hydrodynamical simulations. 
We investigate under what conditions the perfect-merging assumption is appropriate, and infer the planets’ final masses, obliquities and rotation periods, as well as the presence of proto-satellite disks.
We find that the total bound mass and obliquities of the planets formed in our simulations generally agree with $N$-body simulations therefore validating the perfect-merging assumption. 
The inferred obliquities, however, are typically different from those of Uranus and Neptune, and can be roughly matched only in a few cases. In addition, we find that in most cases the planets formed in this scenario rotate faster than Uranus and Neptune, close to break-up speed, and have massive disks.  We therefore conclude that forming Uranus and Neptune in this scenario is challenging, and further research is required. We suggest that future planet formation models should aim to explain the various physical properties of the planets such as their masses, compositions, obliquities, rotation rates and satellite systems. 
\end{abstract}


\section{Introduction}

Uranus and Neptune are the two outermost planets in the solar system. 
Interior models that fit their measured gravitational fields suggest that they are mostly composed of heavy elements (silicates and volatiles) and have  hydrogen-helium (hereafter H-He) envelopes of $\sim 2-3$~\ME (e.g., \citealt{Nettelmann2013, Helled2011}). The latter suggests that their formation timescale must be shorter or comparable to the lifetime of the solar nebula. 
It is commonly assumed that both planets formed via the core accretion model \citep{1996Icar..124...62P}
as ``failed" giant planets:
they form a heavy-element core onto which gas is slowly accreted, but never reach runaway gas accretion \citep{Dodson-Robinson2010,Helled2014}. 
Explaining the formation of Uranus and Neptune is challenging. In classical core accretion simulations, at their current orbital locations, the expected low solid surface density of planetesimals typically leads to formation timescales that exceed the disk's lifetime (\citealt{1969edo..book.....S,1996Icar..124...62P}). Shorter formation timescales could be a result of more massive disks, original formation locations  that are closer to the sun or  higher accretion rates of solids that can be in the form of small planetesimals and/or pebbles (e.g., \citealt{Levison2010,Lambrechts2014,Helled2014}). While such modifications to the model's assumptions can lead to shorter formation timescales, each model suffers from some limitations and the challenge to infer the correct final mass and composition remains \cite[see][for details]{Helled2020}.

The mass ratio of Uranus and Neptune is only 1.18. Explaining their almost unitary mass ratio has been a problem in simulations modelling their growth via pure planetesimal accretion \citep[e.g.,][]{Levison2010}.
Another important constraint on formation models is given by the large obliquities of Uranus  (97\degree) and Neptune (30\degree).
Planets that formed by accretion of small bodies like planetesimals or pebbles are expected to have an obliquity close to zero
\citep{DonesTremaine93}. Since both ice giants have large obliquities it 
implies that both of them experienced at least one major collision involving a massive body during or after their formation \citep{Stevenson1986,Slattery1992,Podolak2012}. 
This motivated \cite{Izidoro2015} (hereafter I15) to propose an alternative formation scenario where Jupiter and Saturn have formed early and served as  a dynamical barrier for smaller planetary embryos (with masses of a few M$_{\oplus}$) that formed at larger radial distances. 
While migrating inwards, these embryos were gravitationally scattered and merged in a series of giant impacts to form Uranus and Neptune. The simulations 
that best reproduce their mass ratio typically begin with five to ten planetary embryos of 3 to 9~\ME, although simulations where the planetary embryos start with masses between 1 and 10~\ME were also relatively  successful \citep{Izidoro2015}. 

This formation path is promising because the formation of planetary embryos in this mass range is more probable. In addition, this model can explain the ice giants' observed obliquities \citep{Slattery1992,Leeetal2007,Morbidelli2012} without the requirement for post-formation giant impacts as investigated in various studies \citep{Kegerreis2018,Kurosaki2019,Kegerreis2019,Reinhardt2020}. 

In the I15 scenario, the embryos were assumed to have merged perfectly after each collision. However, this simplified  assumption cannot capture the details of collisional outcomes such as
hit-and-run events, fragmentation and erosion (target/impactor), the formation of disks, and the effect on the composition and interiors of the planets. 
High resolution 3D hydrodynamical simulations are ideal to investigate these aspects but prior studies assume rather arbitrary initial conditions that are not specifically linked to a suggested formation path. 

In this study we present a suite of hydrodynamical simulations where we track the collision history of Uranus and Neptune analogues as inferred by I15. 
This allows to connect impact simulations with formation models. We investigate the post-impact properties of the final planets, in particular the final mass, rotation period and obliquity, and formation of proto-satellite disks. The paper is organized as follows. In Section\,\ref{section:methods}, we describe how the proto-planets are built and the initial conditions setup for the collisions. In Section\,\ref{section:results}, we present the results which are discussed further in Section~\ref{section:discussion}.  Our conclusions are summarized in Section~\ref{section:conclusions}. 

\section{Methods}
\label{section:methods}
We use the Smoothed Particle Hydrodynamics method (SPH)
%(e.g., \citealt{Monaghan1992}) \commentjs{I think using the Monaghan reference to SPH is to vague and just saying SPH code GASOLINE with references to Wadsley and Reinhardt would be better here. You have it below so you can probably leave it as hydrodynamic simulations of the collisions.}
which was used in prior works for detailed investigations of giant impacts (e.g., \citealt{Canup2001a}, \citealt{Genda2012}, \citealt{Emsenhuber2018}, \citealt{Chau2018}, \citealt{Kegerreis2018}, \citealt{Kegerreis2019}, \citealt{Kurosaki2019}, \citealt{Reinhardt2020}).
Each collision is drawn from the $N$-body simulations of I15. The SPH initial conditions for these collisions are set up following \cite{Reinhardt2020} and simulations are performed using the SPH code \textsc{Gasoline} \citep{Wadsley2004} with the modifications for planetary collisions described in \citet{Reinhardt2017,Reinhardt2020}. 
The simulations were run until the mass of the final planet-disk system converged (which was usually around $t=$27~hours post collision).

\subsection{Equilibrium models \label{subsection:equilibriummodels}}
The particle representation of the colliding embryos is generated with the code \textsc{ballic} (see  \citealt{Reinhardt2017,Reinhardt2020} for details).
The initial thermal state and composition of the embryos 
are not determined in I15. Since the embryos are assumed to have formed beyond Saturn's orbit,
i.e., beyond the water snow line, they are 
likely to contain a large amount of water ice.

The initial masses of the embryos vary between 1 and 9~\ME. As a result, it is unlikely that they accreted a significant amount of H-He prior to the collisions \citep[e.g.,][]{Helled2014}. We therefore neglect the H-He envelope and model the initial bodies assuming a silicate core and an ice mantle, with  ice-to-rock ratio of 1:1. 
We use the Tillotson equation of state \citep{Tillotson1962} to model the heavy elements: granite  \citep{Benz1986} for the silicates and ice \citep{Benz1999} for water-ice (H$_2$O).
In order to investigate the sensitivity of the results to the initial composition, we also performed a suite of simulations assuming rock-to-ice ratios of 1:0 (undifferentiated), 7:3 and 3:7. 

In some cases the time between two collisions is between 1 to 2 Myr (as discussed below) which could allow the accretion of a H-He envelope onto the protoplanet during this time. Therefore,  we also add a H-He layer (of 0.5, 1 and 2~\ME) to the planets in order to investigate its effect on the results. 

\begin{figure}
	\centering
	\includegraphics[keepaspectratio,width=0.45\textwidth]{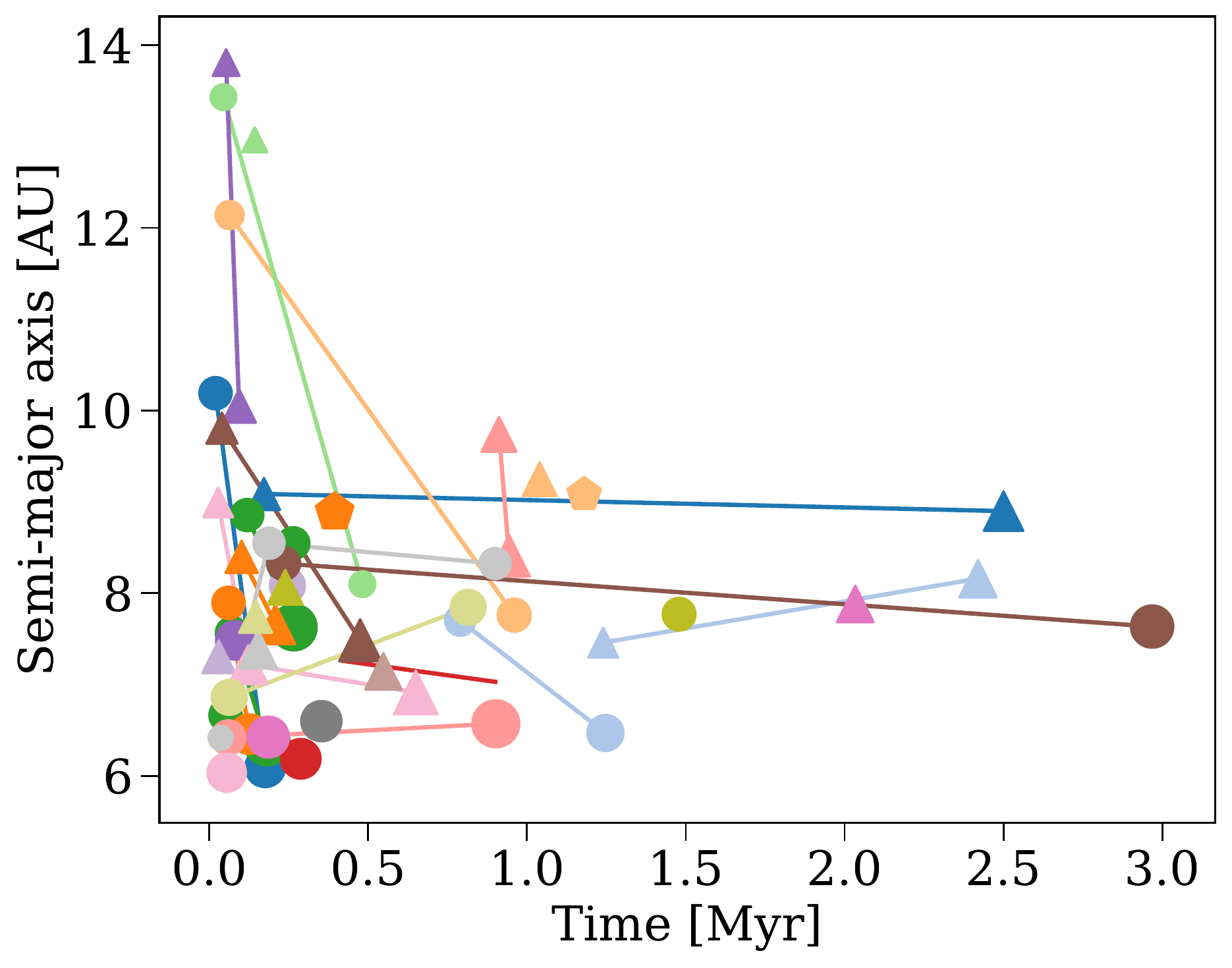}
	\caption{\textbf{The location (semi-major axis) and time (from the start of I15's simulations) of each collision}.
      Every data point corresponds to a collision so each line represents the merging history of a planet. The symbols (circle, triangle, pentagon) distinguish the two or three final planets and each color represents a set of simulations.
      The initial conditions are taken from the $N$-body simulations described in I15. Most of the collisions occur within 0.5 Myr and hence the embryos are not expected to accrete a  substantial amount of H or He between collisions. However, in some cases (e.g., brown circle or blue triangle) more than 2 Myr pass between two collisions. Since these bodies already contain more than 10~\ME of heavy elements they might accrete a substantial H-He envelope prior to the last collision (see Section\,\ref{subsection:equilibriummodels} for details).
  	}
	\label{fig:semimajoraxis_vs_time}
\end{figure}

\begin{figure}
	\centering
	\includegraphics[keepaspectratio,width=0.45\textwidth]{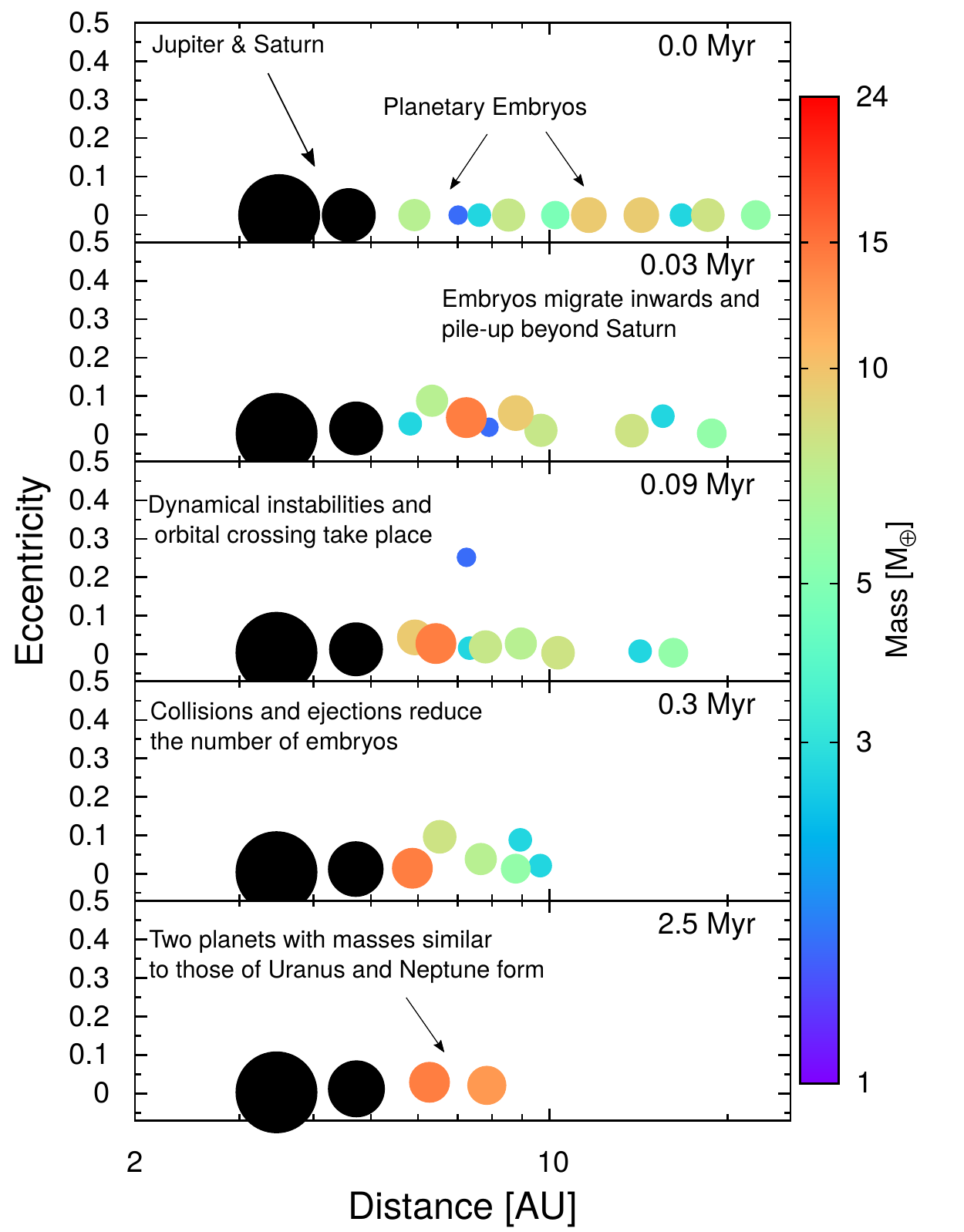}
    \caption{\textbf{Accretion of Uranus and Neptune from inward-migrating massive embryos.}
    The simulations correspond to the study of I15 where the simulations begin with fully formed Jupiter and Saturn orbiting at 3.5 AU and 4.58 AU, respectively.
    The gas giants are a dynamical barrier for a system of smaller, similar-sized planetary embryos that formed at larger radial distances and migrate inwards due to interactions with the gaseous disk. The  embryos then pile-up beyond Saturn and merge in a series of collisions or are ejected from the Solar system. A successful simulation produces two planets with masses similar to Uranus and Neptune beyond Saturn's orbit.
    }
	\label{fig:scenario}
\end{figure}

\subsection{From $N$-body simulations to hydrodynamic initial conditions\label{methods:nbody_to_ic}}

I15 used $N$-body numerical simulations to model the formation of Uranus and Neptune including the effects of a gaseous protoplanetary disk (see Figure \ref{fig:scenario}). Their simulations assume, from the beginning, fully formed Jupiter and Saturn and a population of planetary embryos beyond the orbit of Saturn. Planetary embryos gravitationally interact with the gas disk and migrate inwards. This phase of migration promotes close-encounters and collisions among protoplanetary embryos that grow to masses comparable to those of Uranus and Neptune.  I15 performed a large number of simulations testing the effects of different  initial number  and masses of the population of planetary embryos initially beyond Saturn. A simulation is considered successful in reproducing Uranus and Neptune if the final planetary system contains at least two planets beyond Saturn with masses larger than 12 M$_{\oplus}$. In addition, the mass ratio between the largest ice giant analogues is required to be between 1 and 1.5, and each of them have to have experienced at least one giant collision. In order to also be consistent with the structure of the inner solar system, I15 reject simulations where one or more planetary embryos, that are initially beyond Saturn, invade and get implanted into the inner solar system. They refer to these planets as ``jumpers''. 

The radius of each embryo is computed from its mass and an assumed mean density of 3 g cm$^{-3}$.
A collision between two bodies is detected if their distance is smaller than or equal to their mutual radii.
When creating SPH initial conditions, we evolve the $N$-body collisions in isolation
until the two bodies are separated by 1.2 mutual radii. These modified positions and velocities are then transformed into the common center-of-mass frame and used as initial position and velocity in our impact simulations.

Here we select simulations from three scenarios of I15 that produced high fractions (>~10\%) of Uranus-Neptune analogues. In the first scenario, simulations start with  10 planetary embryos of 6 Earth masses each. In the second and third scenarios the masses of planetary embryos are randomly selected  between 3 and 6, and between 1 and 10 Earth masses, respectively. In all these scenarios the initial total mass in planetary embryos in each simulation is about 60 Earth masses. Planetary embryos are initially randomly distributed past the orbit of Saturn separated from each other by 5 to 10 mutual Hill radii (e.g., \citealt{Kokubo2000}) and released to migrate inwards (for more details see I15).

\subsection{Treatment of the post-impact embryo for the subsequent collision \label{methods:target_internal_state}}
In our set of simulations, most merging embryos experience more than one collision.
This requires some assumptions on the evolution of the thermal state of the   mergers between two impacts. 
The latter can play a role on the final basic properties of the planets such as the mass (e.g., \citet{Chau2018}). In addition, when a circumplanetary disk forms around the merger, the disk structure could substantially change over timescales of $10^4$ and $10^6$ years \citep{2020arXiv200313582I}. 

Since the precise thermal state of the planet and the orbiting material over such long timescales is unknown, and its determination is beyond the scope of this paper, we consider two end-member cases. In the first, we assume that the planet has completely relaxed  and accreted the orbiting material. We therefore build a pristine body with \textsc{ballic} assuming a similar surface temperature as the pre-impact embryos (see \citealt{Chau2018}) and assign it the rotation period and obliquity obtained in the prior collision. In the remaining of the paper, we refer to this case as the "compact/cold" case. In the second case, we directly use the post-impact body as an input for the subsequent collision. We refer to this case as the "expanded/hot" case.

\subsection{Analysis\label{subsection:analysis}}
The analysis of the simulation output is similar to the one presented in \citet{Reinhardt2020}. 
The mass of the gravitationally bound material after a collision is determined using \texttt{SKID}\footnote{The source code is available at: \url{http://faculty.washington.edu/trq/hpcc/tools/skid.html}.} \citep{Stadel2001}
with similar parameters ($nSmooth=1600$ and $tau=0.06$).

We refer to the central part with densities above the uncompressed density of ice, $\rho_0$=0.917 g cm$^{-3}$, as \textit{planet}, which is surrounded by an envelope of gravitationally bound, low-density material. This orbiting material can be further divided into an extended \textit{atmosphere} and a \textit{circumplanetary disk}. 
The disk's mass is estimated using the algorithm described in \cite{Canup2000}. 
First, the particles that belong to the planet are found using a given mean density. Since the mean density of the final planet in our simulations is a priori unknown, we use  $\rho=1.3$~g\,cm$^{-3}$ which is similar to Uranus' mean density. 
Because the mean densities of the initial embryos in our simulations are higher, we tend to overestimate the planet's size. 
This in turn affects the distribution of the material between the planet and the disk and as a result,  
the inferred disk masses should be taken as lower bounds. 
Then the disk finder computes the keplerian orbit from a particle's angular momentum and orbital energy. If the periapsis is smaller than the planet's radius, the particle's mass is added to the planet. Otherwise it is assigned to the disk. The procedure is repeated iteratively until the final masses converge.

In \cite{Reinhardt2020}, all the collisions were set up to occur in the x-y plane. Since the initial conditions in this paper are obtained from $N$-body simulations, the analysis is more challenging because the collision plane is usually not aligned with any of the coordinate axes. We therefore compute the planetary  rotation period $P$ (solid-body) from the particles' angular momentum and moment of inertia solving
\begin{equation}
    \vec{L} = I \vec{\omega},
\end{equation}
 for $\vec{\omega}$ the angular velocity vector. The orientation of the rotation axis is given by $\vec{\omega}/\norm{\vec{\omega}}$, and the rotation period by $P=2\pi/\norm{\vec{\omega}}$.
The planet's obliquity is obtained from the angle between the (normalized) angular velocity vector and a unit vector that is perpendicular to the x-y plane which corresponds to the solar plane in I15's simulations.

\section{Results}
\label{section:results}
All collisions occur at impact velocities below the mutual escape velocity.
This means that bodies merge and little mass is lost in a collision. Usually the colliding embryos are assumed to merge in the first encounter (hereafter, \textit{direct mergers}). 
About half of the cases we investigated do not lead to a direct merging of the two colliding embryos. Instead, they bounce off each other a few times while remaining gravitionally bound (hereafter, \textit{bouncing  mergers}). The colliding embryos hence experience several encounters, where each impact has a different impact velocity and angle. The bodies are typically tidally deformed during each encounter. In some extreme cases one of the bodies (or both) is tidally stripped and the material forms arm-like or spiral-like structures. This leads to diverse final outcomes.   

\begin{figure*}
	\centering
	\includegraphics[keepaspectratio,width=\textwidth]{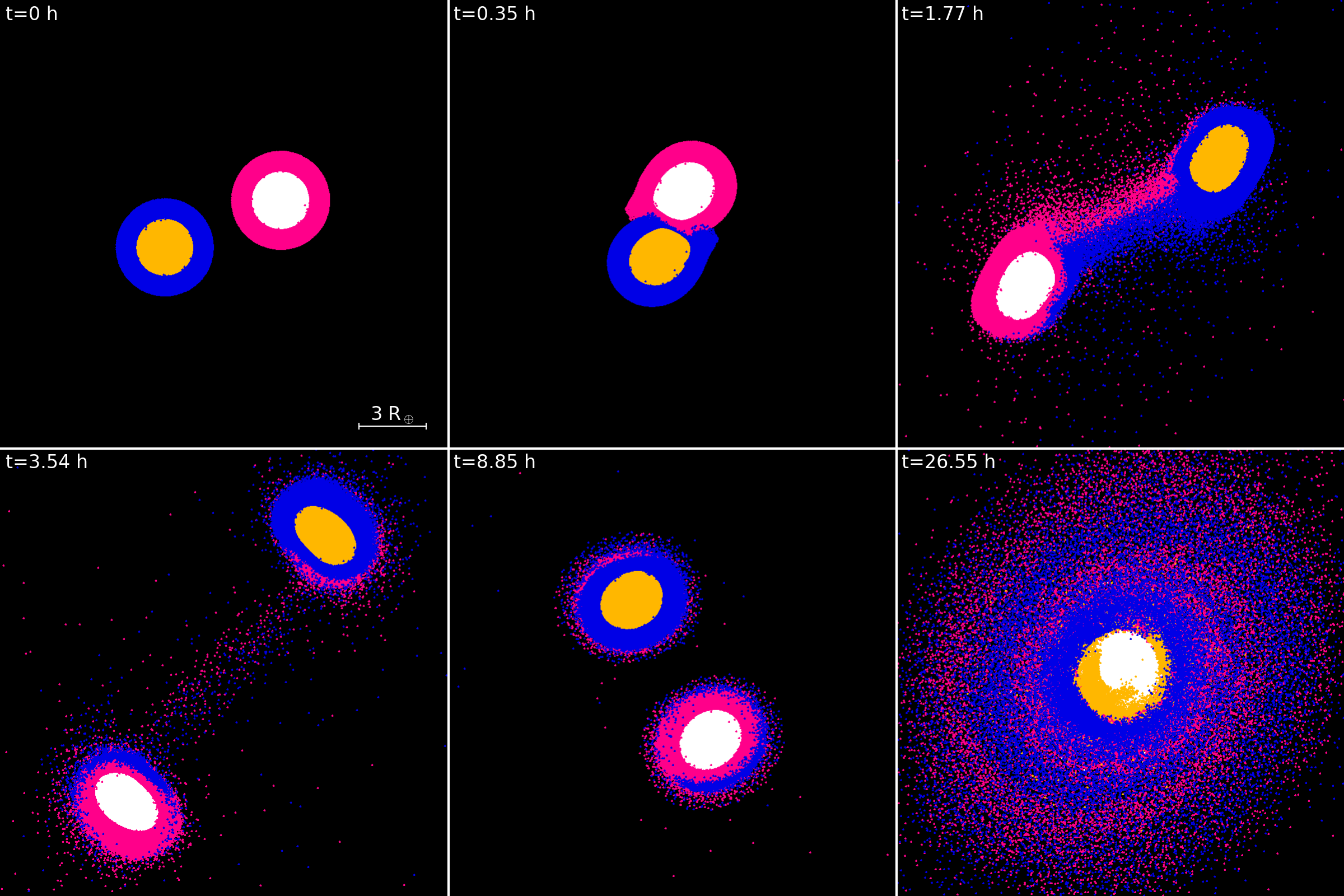}
    \caption{\textbf{Snapshots of a bouncing merger (G11c).} The two bodies of 9.48~\ME and 9.10~\ME collide grazingly at first contact and continue on their (perturbed) orbits. They then recollide and merge at the second contact. After 27 hours, the system settles in a central body of 18.57~\ME   surrounded by a massive disk of 1.413~~\ME. The colors identify the different materials and their origin: yellow and white for rock, blue and magenta for water-ice.
    }
	\label{fig:bouncingmerger_p1_130}
\end{figure*}

When the embryos are similar in mass (within 10\%), they tend to merge after a few encounters and form a central body with a massive disk, similar to the direct mergers. By symmetry of the collision the material of both bodies is usually well mixed. This occurred in 33 out of 39 "bouncing mergers".
One example is shown in Figure\,\ref{fig:bouncingmerger_p1_130}.

When the embryos differ in mass or collide at large impact angles it is possible that one
embryo is tidally stripped during the first encounter. The stripped material forms a very long tail 
that fragments into smaller clumps. The central part of the disrupted body then recollides with the remaining embryo. 
Most simulations show a tail wrapping the central body, which is more likely to put the clumps on bound orbits. The clumps are then either tidally disrupted or merge with the planet. Some clumps also merge with each other and form larger bodies.
In a few cases these clumps remain in orbit during the whole simulation (two weeks).
Only in one case, a long straight tail is formed, where the clumps are rather put on unbound orbits which leave the system.
In this singular case, a few remaining clumps collide with the central planet and the system converges in a few days without forming a massive disk. 
    
\begin{figure*}
	\centering
	\includegraphics[keepaspectratio,width=\textwidth]{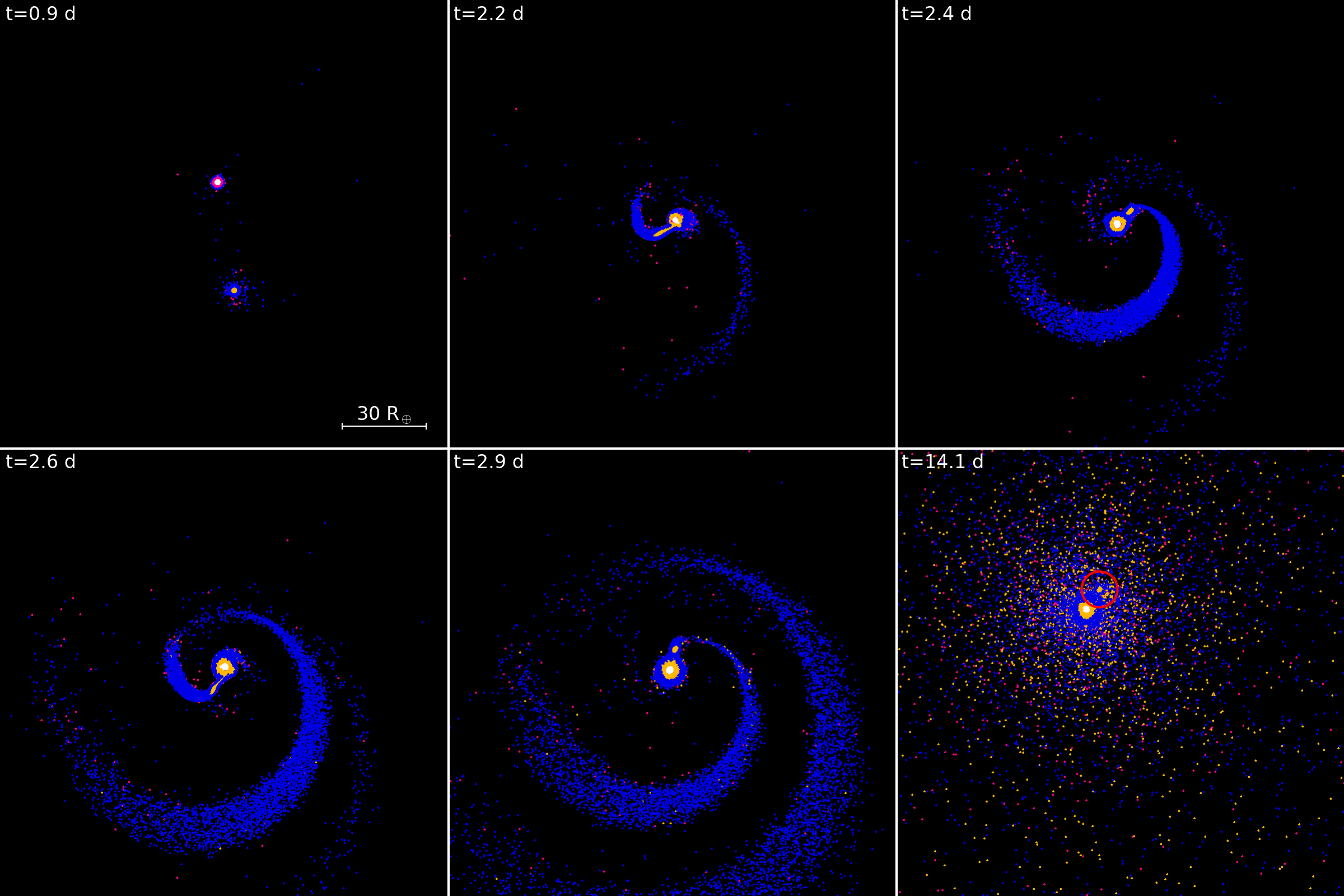}
    \caption{\textbf{Snapshots of a non-merging system (H21c).} The two embryos of 7.37~\ME and 4.64~\ME first experience a grazing impact and continue on their (perturbed) orbits but remain gravitationally bound and collide several times before finally merging. 
    At every contact the smaller body is tidally disrupted and moves in closer until the system reaches a (pseudo-)equilibrium state. The orbiting impactor remnant (marked in red in the last snapshot) remains in orbit for the rest of the simulation. The colors identify the different  materials and their origin: yellow and white for rock, blue and magenta for water-ice.
	\label{fig:nongmerger_p1_242}}
\end{figure*}

In five cases (with mass ratios between 1:2 and 1:5) an intermediate scenario occurs. The two bodies experience several encounters that successively strip the mantle and part of the core from both embryos
and result in a spiral of ejecta. We observe that the smaller body's core remains close to the Roche lobe of the larger one. After an initially rapid mass loss it enters a state where tidal deformation and self-gravity seem be in a (potential unstable) equilibrium and remains on this orbit for the rest of the simulation (up to two weeks). An example of such a collision is presented in Figure\,\ref{fig:nongmerger_p1_242}. 

\subsection{Obliquities}
\label{subsection:obliquities}
Figure\,\ref{fig:obliquity_vs_mass} shows the obliquity's evolution as well as the final planets' obliquities for both types of merging impacts.
The inferred planetary obliquity is in most cases either a few degrees or close to 180. 
Consequently, only a few cases are close to the current obliquities of Uranus and Neptune (97 and 30 degrees, respectively) 
and might be considered as analogs. 
No simulation manages to produce a Uranus and a Neptune analogue in obliquity simultaneously. However, two sets of simulations result in two final planets with obliquities close to Uranus and Neptune (orange and brown lines in Figure\,\ref{fig:obliquity_vs_mass}).
Since direct mergers are well described by perfectly inelastic collisions these results are in good agreement with the obliquities found in I15 that were calculated from  angular momentum conservation.
The obliquities resulting from bouncing mergers are similar to the ones found in the direct merger cases because the colliding embryos remain in the initial collision plane.

\subsection{Rotation periods}
\begin{figure}
	\centering
	\includegraphics[keepaspectratio,width=0.45\textwidth]{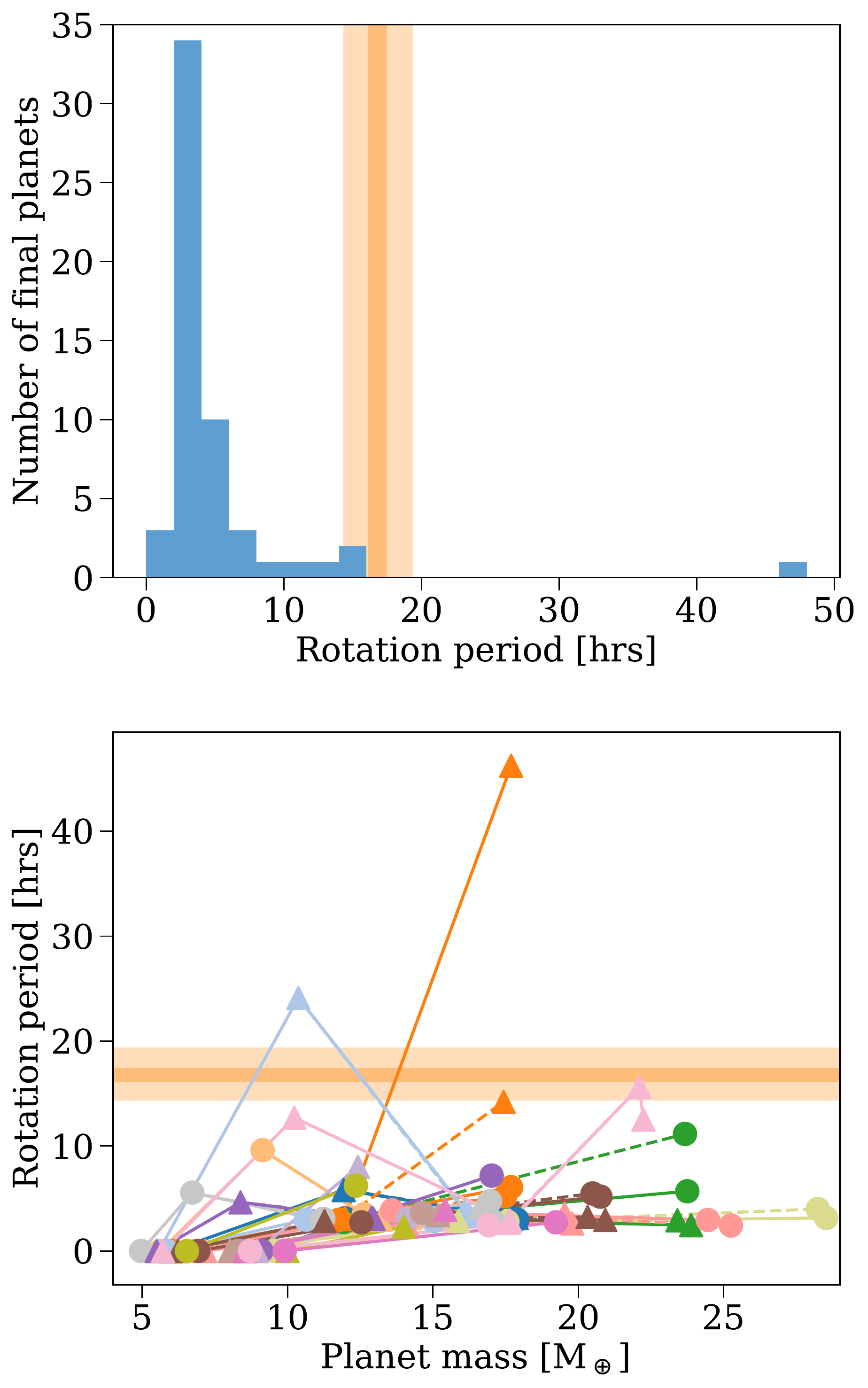}
    \caption{
    \textbf{The planet's rotation period.}
    In both panels the dark orange line represents values that are consistent with Uranus' and Neptune's inferred rotation periods and the light orange bar indicates values that agree with the mean value within 15\%.
    Each planet's rotation period is inferred from its angular momentum and moment of inertia values after the simulations converged (which is usually 27 hours after the impact) as described in Section\,\ref{subsection:analysis}. The few collisions that have not converged after two weeks of in simulation time are not included in the plot.
    \textbf{Top panel:}
    The inferred distribution of the planet's final rotation periods. Most planets rotate extremely fast with $P$ being $\sim$ 3-5 hours. A few planets rotate slower and only two (but from different simulation sets) have rotation periods that are comparable to Uranus or Neptune.
    \textbf{Bottom panel:}
    Each color represents a set of simulations, while the symbols distinguish the two or three planets. The lines indicate the assumed thermal state of the embryo that experienced a prior collision, thick line for a "compact/cold" and dashed for a "expanded/hot" initial state. Usually the extreme rotation periods are already obtained in the first collision an embryo experiences and subsequent impacts have little effect on the final value (one impressive exception being the orange triangle). 
    	\label{fig:period_vs_mass}
    }
\end{figure}

In our simulations we also determine the planets' rotation period as described in Section\,\ref{subsection:analysis}.  Figure\,\ref{fig:period_vs_mass} shows the rotation period after each collision and the final rotation period. 

The final periods are extremely short, with $P$ being $\sim$ 3-5 hours, close to the rotational break-up speed of approximately $\sim$ 1.9 hours. 
The fast rotation periods, which are close to the ones expected from angular momentum conservation, are often already obtained in the first collision.
This can be explained by the initial mass distribution of the embryos, where most embryos have similar masses, since for a given colliding mass equal mass mergers contain more angular momentum than a higher mass ratio (see Appendix \ref{appendix:angularmomentum}). 

We find that often the subsequent collisions do not substantially alter the rotation period but rather change the orientation of the planet's rotation axis.
A significant change is only observed when the angular momentum vectors are (almost) aligned, i.e., if the spinning bodies rotate in the same plane as the collision plane. In that case, the rotation period can either be decreased if the two vectors are aligned, or increased if the two vectors are anti-aligned.
In the most extreme case observed in our simulations, the rotation period is increased by a factor of six, while a few intermediate cases double their rotation period in the final collision.
Only in a few simulations (orange triangle, pink triangle, green circle) do the final planets have rotation rates that are similar to observed rotation periods of Uranus and Neptune which are of the order of 15-17 hr \citep{Helled2010}. 
Again, we do not find any Uranus and Neptune analogues in rotation period simultaneously in the simulations. 

\subsection{Formation of disks \label{subsection:disks}}

\begin{figure}
	\centering
	\includegraphics[keepaspectratio,width=0.45\textwidth]{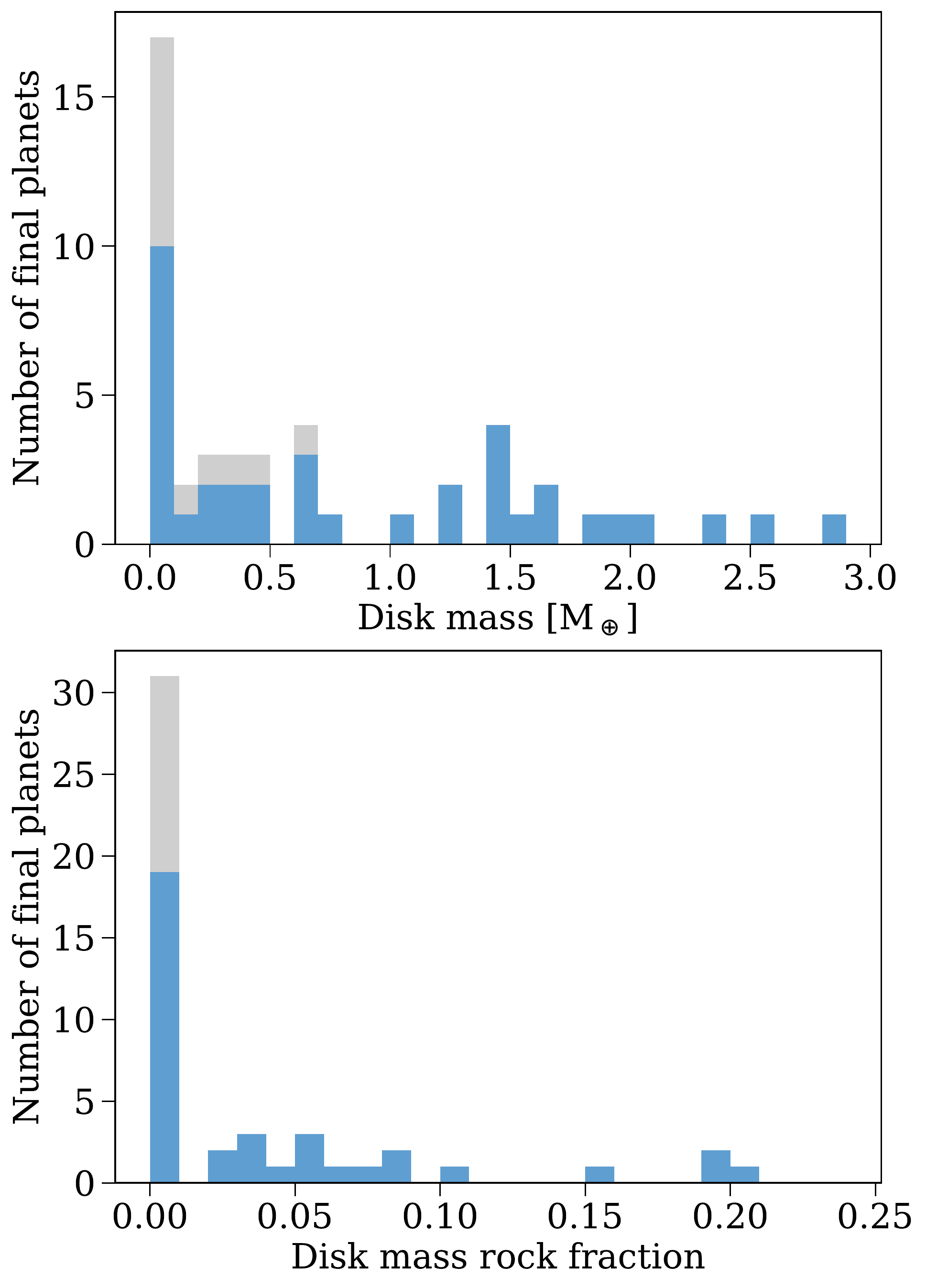}
    \caption{{\textbf{Final disk mass and rock mass fraction.} The color distinguishes what type of collisions a planet experienced (grey: only direct merger, blue: at least one bouncing merger).
    \textbf{Top panel:} The top panel shows the distribution of the disk masses inferred from our simulations. If a growing planet only experienced direct merger the resulting disks contain less than 1~\ME. If one (or more) bouncing mergers occurred in a planet's accretion history the disks are much more massive with masses between 0.5 and  2.7~\ME.
    \textbf{Bottom panel:} The distribution of the disk rock mass fraction inferred from our simulations. Disks around planets that only grew via direct mergers contain little rock (but usually still 10-100 the rock mass in all five major Uranian satellites). If a planet experienced at least one bouncing merger the resulting disk can contain up to 20\% of rock. As shown in Figure \ref{fig:diskrockmass_vs_mass} these extreme cases are also the most massive disks found in our simulations.
    }}
	\label{fig:disk_mass_hist}
\end{figure}

\begin{figure}
	\centering
	\includegraphics[keepaspectratio,width=0.45\textwidth]{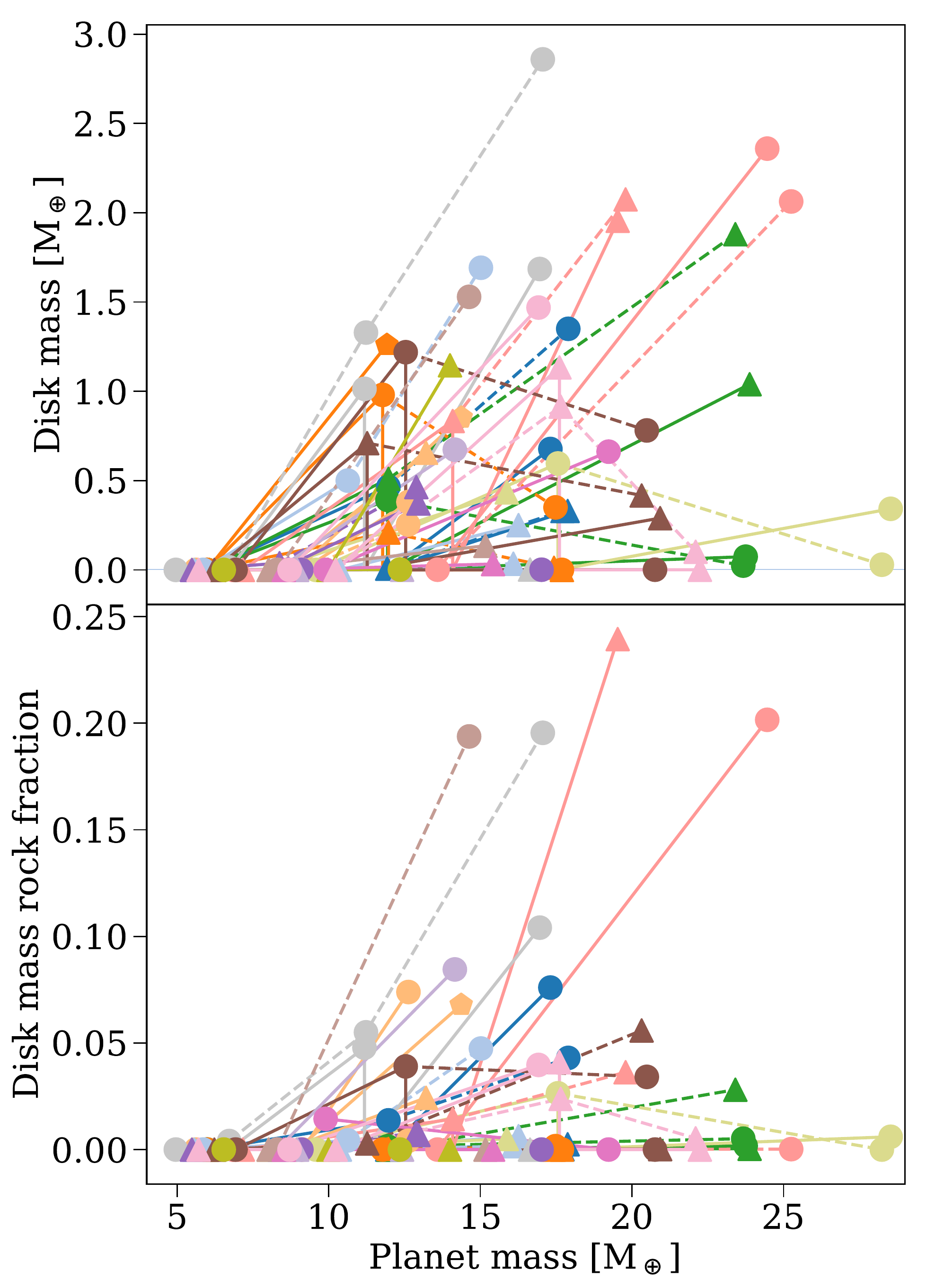}
    	\caption{
    	\textbf{The disk mass after each collision and the evolution of the silicate mass in the disk.}
    	Each color represents a set of simulations, the symbol distinguishes the two or three final planets that form and each data point along a line represents the disk mass and rock mass fraction in the disk respectively after the respective collision, the last data point therefore corresponds to the final disk mass and silicate mass fraction in the disk. 
    	The disk mass is inferred once the simulation converged (which is usually 23 hours after the impact) as described in Section\,\ref{subsection:analysis}. In many cases the resulting disk is very massive and can contain up to 2.9~\ME (or 17\% of the total colliding mass). This is several orders of magnitude larger than the mass of Uranus' regular  satellites.  The disks that are more massive than 0.5~\ME are all obtained in "bouncing mergers" (see 
    	Section\,\ref{subsection:bouncingmergers} for details). Subsequent collisions often increase the disk mass but can also erode (e.g., brown circle) or even disrupt (e.g., light green circle) an existing disk.
    	In most collisions the disk is enriched in rock and contains several times the amount of rock needed to form Uranus' five regular satellites. In the most extreme case (grey circle), which also produced the most massive disk we observed in our simulations %(see Figure \ref{fig:diskmass_vs_mass}) 
    	the disk contains 20\% of rock.
    	}
	\label{fig:diskrockmass_vs_mass}
\end{figure}

We find that most of the collisions lead to the formation of a circum-planetary disk, with their mass and composition  depending on the type of mergers. For direct mergers, the disk is usually massive and contains between 0.1 and 0.5~\ME~or 1 to 5\% of the colliding mass (see Figure\,\ref{fig:disk_mass_hist}).
The disks are mostly composed of water but also contain some rock, see Figure\,\ref{fig:diskrockmass_vs_mass}. The fraction of rocks is up to 2\% of the disk's mass, or about 10$^{-2}$~\ME. 
For the ``expanded/hot" cases, where we keep the disk generated in a previous impact, the disk can increase in mass, be partially eroded or even destroyed in the second collision.
If the disk survives we observe that in most subsequent collisions it tends to be enriched in rock, i.e., the total mass of orbiting rock increases.
The final amount of rock in the disk varies between 10$^{-4}$ and 10$^{-2}$~\ME. These later cases contain then several times the total rock mass of Uranus' major five satellites ($0.75 \times 10^{-3}$~\ME~where $M_S=1.5 \times 10^{-3}$~\ME~is the total satellite mass).
The well defined disks are aligned with the planet's rotating plane $\pm~1$ degree.

For the bouncing mergers the disks are found to be very massive where over 50\% of them contain between 0.5~\ME~and 2.9~\ME~(5--17\% of the total bound mass).
We also observe that the disks are substantially enriched in rock with the rock mass fraction being up to 15\%. 
Compared to the direct merger cases, where the disk mass does not exceed 0.5~\ME, these bouncing mergers produce disks that are substantially more massive, more extended, and more enriched in rock.

\begin{figure}
	\centering
	\includegraphics[keepaspectratio,width=0.45\textwidth]{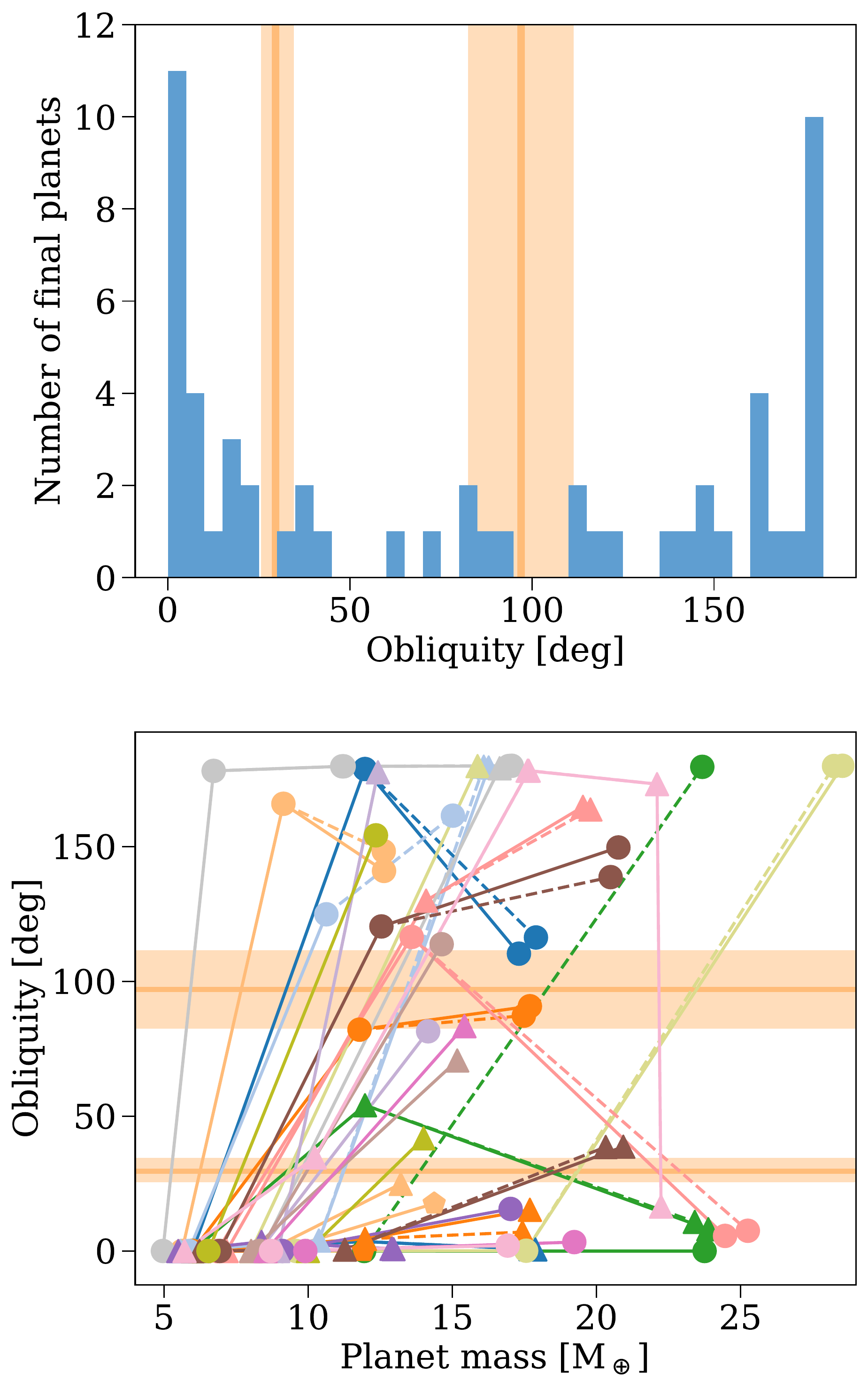}
    	\caption{
    	%The planet's obliquites versus mass (top) and final obliquities (bottom).
    	\textbf{The planets' obliquities.}
    	In both panels the dark orange line represents Uranus' and Neptune's obliquities and the light orange bar indicates values that agree within 15\%.
        Each planet's obliquity is inferred from its angular momentum and moment of inertia values after the simulations converged (which is usually 27 hours after the impact) as described in Section\,\ref{subsection:analysis}. The few collisions that have not converged after two weeks of in simulation time are not included in the plot.
        \textbf{Top panel:}
        The distribution of the planet's final obliquities inferred from the  simulations. Most planets have values that are either close to zero or 180 degrees.
        \textbf{Bottom panel:}
        Each data point along a line represents the planet's obliquity after the respective collision. Since the planets' mass increases with each collision the first data point (at the lowest mass) corresponds to the embryo's initial value and the last data point (at the largest mass) to the final planetary obliquity. Each color represents a set of simulations and the symbols (circle, triangle, pentagon) distinguishes the two or three final planets. The lines' style indicate the assumed thermal state of the main embryo (described in Section\,\ref{methods:target_internal_state}), thick line for compact and dashed for expanded. The obliquity can change substantially due to each collision and the final value is determined by the last collision a planet experienced.}
    		\label{fig:obliquity_vs_mass}
\end{figure}

% Final period evolution

\subsection{Sensitivity to the composition and thermal state of the colliding bodies \label{subsection:sensitivity_target}}
The results presented above were derived for the canonical composition assuming an ice-to-rock ratio of 1:1. In order to explore the sensitivity of these results to the composition we performed several impact simulations with different ice-to-rock ratios. For the few simulations that had a long time interval in between collisions we also considered the accretion of a H-He envelope (see Section\,\ref{subsection:equilibriummodels} for details).

We find that the final planetary masses are rather  insensitive to the assumed composition, while the rotation periods are only weakly sensitive to the assumed composition (with a variation of $\sim$ 5\%). 
The largest differences are found in the planet-to-disk mass ratio and the disk composition.
We find that the mass of the disks varies on the order of 20\% with no clear trend, while the more rock the embryos contain, i.e. the lower the ice-to-rock ratio, the more rock is incorporated into the disk.
For models with a H-He layer, we find that the disks are smaller and contain a substantial mass fraction (30 to 70\%) of H-He. The heavier the envelope of the planet, the more gas is captured into the disk. 

Another assumption is the thermal state of a planet that experienced a previous collision. 
Comparing the collision outcomes of these "condensed/cold" cases to the "expanded/hot" ones discussed in Section\,\ref{methods:target_internal_state}
we find good overall agreement.
The planets' masses as well as their obliquities are very similar.
Initially compact objects tend to rotate slightly faster (10-15 \%) than their extended counterparts because the impactor material and angular momentum is deposited in the upper part of the planet (in the compact case). Vice versa in the extended case, the outer layers of the "planet" are low density and the impactor penetrates deeper and deposits angular momentum in the deeper part of the planet.
Only in two/three cases is the obliquity/rotation period significantly different from one assumption to the other. 

The differences are most pronounced 
in the evolution of the disks, in particular, whether or not a preexisting disk is considered. Figure\,\ref{fig:disk_evolution_all} shows the mass and obliquity evolution for the cases that form massive disks. 
The exact outcome in mass depends on the specifics of the collisions.
In some suites of simulations, a large part of the first disk survives and merges with the new disk (e.g., planet 2 in simulation a), 1 in e), 1 in g), 2 in h) or 1 in i)). The corresponding simulations with compact initial embryos lead to the formation of massive disks ($>0.5~M_{\oplus}$).

On the contrary, impact conditions that do not form massive disks in the compact case
result in the partial or total destruction of a pre-existing disk in the extended case (e.g. 1 in c), 1 in d), 2 in f) and 1 in h)).

The disk obliquities follow closely the planets' ones, as expected from angular momentum conservation. 
The obliquity evolution is little sensitive to the assumed thermal state. A few extended cases exhibit a slight misalignement of a few degrees compared to the compact cases (e.g., 1 in d), 1 in e), 2 in h)). This could be explained by some disk particles that are not or little perturbed by the collision, hence stay close to their previous orbits and contribute to the misalignement of the final disk. 

Generally, the final disk's mass and orientation are predominantly determined by the last collision. 
However, under some favorable circumstances pre-existing disks can survive and contribute to the final disk.

The initial extent of the planet can also alter the collision evolution as some cases enter the bouncing regime in the compact cases while they do not in the extended case. This has also consequences for the composition. In two extremes cases, the compact cases lead to the fragmentation of the tail. This creates more clumps that end up in the massive disk and enrich it in rocks, increasing the total mass and the rock mass fraction by one or two orders of magnitude in comparison to their expanded counterparts.
However, the presence of a massive expanded disk can also lead to the impactor stripping more of the "naked" core which eventually also enriches the disk. 

As discussed above, most quantities are only weakly dependent on the different assumptions, however the disks' properties such as their masses, extents and compositions depend strongly on the thermal state and composition of the initial embryos.

\begin{table}
\begin{tabular}{ c  c  c  c  c } 
\toprule
 & \tiny{Obliquity} & \tiny{Rotation period} & \tiny{Disk mass} & \tiny{Disk rock mass} \\
\toprule
1a & x            & $\checkmark$  & $\thicksim$   & x             \\
1b & $\checkmark$ & $\thicksim$   & x             & x             \\ \midrule
2a & x            & $\checkmark$  & $\checkmark$  & $\thicksim$   \\
2b & x            & x             & $\checkmark$  & $\thicksim$   \\ \midrule
3a & x            & x             & $\checkmark$  & $\checkmark$  \\
3b & $\checkmark$ & x             & x             & x             \\ \midrule
4-28a & x            & x             & $\checkmark$  & $\checkmark$  \\
4-28b & x            & x             & x             & x             \\
\toprule
\end{tabular}
	\caption{\textbf{Cases that can lead to Uranus/Neptune analogs.} Each doublet is a set of simulations where at least one of the planets matches Uranus or Neptune in (at least) one of the investigated properties. A successful match in obliquity or rotation period agrees within 15\% with the observed values. Disks are considered a potential proto-satellite disks for the Uranian satellites if the disk mass is at least 10$^2$ $M_S$ (where $M_S$=1.5$\times$10$^{-3}$~\ME is the total mass of the five regular satellites) or when the disk contains at least 50\% of  the satellite mass in rock. None of the simulations produced a planet that simultaneously matches all the investigated properties. }
	%Obtaining a Uranus and a Neptune analogue in all aspects in the same simulation is therefore very unlikely. 
	\label{table:bestcases}
\end{table}

\section{Discussion}
\label{section:discussion}
We follow the collisional history of the massive embryos introduced in I15 that can lead to Uranus-like and Neptune-like planets using 3D hydro simulations. We then investigate the mass, rotation rate, and obliquity as well as the formation of circumplanetary disks in each collision. 
In our simulations, we find two classes of merging impacts: ``direct mergers" where the bodies merge in the first collision and ``bouncing mergers" where the two bodies bounce off each other but remain bound. They usually experience several grazing encounters before merging. While the results are generally similar, we observe very strong tidal deformations of one
or both bodies between encounters in the bouncing collisions. In some cases, we also detect fragmentation of the stripped material into smaller clumps. While most of them are either tidally disrupted or merge with the planet (or with each other), some remain in orbit over the entire simulation (two weeks in simulation time).

In five extreme cases, the remnant of the impactor's core is tidally eroded but remains in orbit around the larger embryo, just beyond the Roche limit. After an initial phase of rapid mass loss it enters an equilibrium state and forms a very massive satellite ($\sim$1.5\% of the planet's mass). It is questionable whether this configuration remains stable over longer time scales and further investigation, e.g., with higher resolution to better resolve the tidal erosion, is required.
Should they persist over larger time scales both the smaller orbiting clumps and the impactor remnant will not match Uranus' or Neptune's current satellite system because they are too massive and too close to the planet. 

In both types of mergers, the final masses and the obliquities generally agree with the results of I15 therefore validating the perfect merging assumption in these cases. 
The planets' final obliquities are determined by the last major collision they experience.
Figure \ref{fig:obliquity_vs_mass} shows that obtaining the desired obliquities is possible but rather  unlikely because most of the inferred values are either close to zero or 180 degree. 
This is probably due to the strong dampening of the planet's orbital inclination in I15's simulations which favours collisions in the orbital plane.
If the planets were allowed to collide with larger inclinations the expected obliquity distribution could have been closer to an isotropic one. In this case the probability to obtain the desired value in a single giant impact is $\sim 1/180$. Clearly, further investigation of this topic is required. 

Finding Uranus Neptune analogues in obliquity in the same simulation is very challenging and requires very specific impact conditions. 
In principle the possibility of another impact with a smaller body like in \citet{Reinhardt2020} cannot be excluded. \citet{Rogoszinski2020} found that the presence of a massive circumplanetary disk produced during gas accretion from the protoplanetary disk can enhance spin-orbit resonances and change the planetary obliquity over several Myrs.
While the accretion of a H-He envelope is not included in our simulations the formation of massive debris disks seems to be robust.
Since these debris disks are more massive and compact than the accretion disks considered in \citet{Rogoszinski2020} their effect on the precession rate of the planet's spin axis is currently difficult to estimate.

In most simulations the forming planets  rotate very fast with P$\sim$3 - 5 hours, close to the break-up velocity.
These extreme values are usually obtained already in the first collision and subsequent collisions do not change the rotation period substantially (as shown in Figure\,\ref{fig:period_vs_mass}), apart from one exception (orange triangle in the same Figure).  
Only two planets (out of 56 produced in our simulations) have final rotation rates that are comparable to those of  Uranus and Neptune. 
Again, we do not find a Uranus and a Neptune analogue in the same simulations. 

This emphasises the challenge in forming Uranus and Neptune.
Although a later impact with a smaller, e.g., 1 to 3~\ME~body, could alter the rotation period (e.g., \citealt{Slattery1992,Kegerreis2018, Reinhardt2020}), in order to substantially slow down the planet very specific impact conditions are needed (e.g., an equatorial impact if the obliquity should be preserved). 
As a result, also in this formation scenario ``fine-tuning" of the impact parameters is needed. The accretion of such a body would also alter the final planet's mass and is likely to also affect its obliquity and interior as well as any existing satellite system (\citealt{Morbidelli2012}, \citealt{Reinhardt2020}). 
Although the obliquity and rotation period of both planets could change during their long-term evolution such a mechanism is yet to be presented. Alternatively, Uranus and Neptune  might have formed from smaller bodies (resulting in a smaller obliquity and rotation period) as suggested by standard formation models that consider planetesimal and/or pebble accretion and subsequently experienced a final, less massive, impact. 

In principle, a massive close-in satellite as found in five bouncing collisions could reduce the planets' rotation period over several Myrs through tidal acceleration as observed in case of the Earth-Moon system. Such a configuration might also result in evection resonances that were proposed in case of the Earth-Moon system as a mechanism to reduce the angular momentum after its formation via a giant impact (\citealt{Cuk2012}).
However, the formation of such a massive orbiting body is rare (5 out of 88 collisions) and it is yet to be determined whether it remains in orbit, merges with the planet or is tidally disrupted.
Also, angular momentum reduction via evection resonances requires very specific conditions and even in case of the Earth's Moon its efficiency is questionable (\citealt{Rufu2020}). 
Such a scenario also introduces another problem: neither Uranus nor Neptune have such a massive regular satellite so it would have to be ejected or disrupted after the planet has slowed down.

We therefore conclude that it is challenging to obtain Uranus' and Neptune's rotation periods through mergers of very massive embryos. This problem does not occur in the standard picture of planetesimal/pebble accretion  followed by a giant impact (post formation). 
However, this scenario also requires rather specific impact conditions and is difficult to reconcile with formation models. 

While the bound mass and obliquity we compute are consistent with I15 we find that the formation of circumplanetary disks is ubiquitous in our simulations.
These disks are typically quite massive (up to 2.9~\ME) and tend to be aligned with the planet's obliquities.
Therefore, a substantial part of the total bound mass remains in orbit and could form satellites rather than being accreted to the  planet. Satellites formed in such disks would be orbiting prograde and close to the planet's equatorial plane consistent with Uranus' major five satellites. Neptune's massive satellite, Triton, is on an irregular, very inclined orbit which is inconsistent with the formation from a circum-planetary disk.
As a result, the presence (and absence) of such a massive disk could not only account for the mass difference of approximately 2.5~\ME between Uranus and Neptune but could also explain the differences in their satellite systems.

Another important aspect  is the disk's composition. The Uranian satellites are estimated to have an ice-to-rock ratio of 1:1.
This suggests that the proto-satellite disk from which they formed contained a substantial amount of rock. Prior simulations that considered smaller (1 to 3~\ME), differentiated projectiles (e.g., \citealt{Slattery1992, Kegerreis2018, Reinhardt2020}) found that the compositions of the impact generated disks are dominated by water ice (and H-He from proto-Uranus' outer envelope), and that it is challenging to eject enough rock to the disk to the explain the moons' composition. The large amounts of rocks in the massive disks obtained in the bouncing cases would naturally result in a rock-rich composition and therefore can explain the composition of the Uranian moons. 

In addition, due to the symmetry of similar mass collisions, the material of the two embryos mixes in comparison to cases where the bodies have very different masses. This affects the final composition of the planet and the proto-satellite disk (if one is formed in the impact).
In the case of Neptune, such a mixed interior could
%result in an adiabatic interior and therefore
explain its larger heat flux and moment of inertia and therefore seems favourable (e.g., \citealt{Podolak2012}). On the other hand, Uranus is in equilibrium with solar insulation and is thought to be more  centrally concentrated, consistent with a stratified interior resulting in inefficient heat transport. 
Such an interior is not a natural outcome of similar mass mergers which result in a well mixed interior.

Finally, our results confirm that most of the planet's properties are dominated by their last collision, reinforcing the importance of giant impacts and collisions in the history of planet formation and evolution. 
However these collisions can show very complex behaviour that are almost chaotic as some assumptions modify the evolution of the collision, which ultimately modifies the planet's properties. 
These behaviours, and especially the formation of complex tails, fragmenting clumps or co-orbiting systems, might be difficult to capture correctly by scaling laws or outcomes of machine learning. For the detailed properties of individual planets, using a full 3D hydrodynamical simulation is required.

\section{Conclusions}
\label{section:conclusions}

Our main conclusions can be summarized as follows:
\begin{itemize}
    \item Perfect merging is a good assumption for most collisions we investigated in this study. As a result, the SPH simulations are typically in agreement with I15. %The merging collisions can be classified in direct and bouncing mergers. A few of the bouncing mergers do not lead to the formation of one final body but rather result in a central planet and a very massive ($\sim$1.5\%) orbiting smaller body.
    
    \item Merging collisions can be classified in direct and bouncing mergers. While the planet's final mass, obliquity, and rotation period are similar for both types of collisions, ``bouncing mergers" lead to the formation of more massive and rock-rich disks.
    
    \item A few of the bouncing mergers do not lead to the formation of one final body but rather result in a central planet and an orbiting smaller body with a mass of $\sim$1.5\% of the planet's mass.
    
    \item The formed planets are found to rotate much faster than the measured rotation periods of Uranus and Neptune. Therefore a post-formation mechanism that significantly reduces the planet's rotation period is required.

    \item The planetary obliquities are typically very different with the ones of Uranus and Neptune (being significantly smaller/larger). However, two simulations leads to two final planets with obliquities comparable to those of the ice giants.
    %with obliquities close to Uranus and Neptune values. 

    \item The formation of disks is  common. The disks are found to be  massive (up to 2.9~\ME) and can contain up to 15\% (or 0.6~\ME) of rock.

    \item No simulation could lead to the formation of two planets that simultaneously agree in obliquity, rotation period and proto-satellite disk with those of Uranus and Neptune.
    
\end{itemize}

It is clear that further investigation of the topic is required. The I15 calculation was based on several simplifying assumptions that should be relaxed in future studies. Better constraints on the embryos original composition and considering gas accretion during migration and between consequent impacts could modify the results and may lead to the formation of planets that are more similar to Uranus and Neptune. 
Also the evolution of the planetary interior and the orbiting material between two collisions must be better modelled and we hope to address these topics in future research. 
Finally, it should be kept in mind that  these properties can change during the planetary long-term evolution, and we suggest that future research should link this  potential formation scenario with the planetary evolution. 
\par

We find that inferring Uranus and Neptune analogues is challenging and requires ``fine tuning".  Our study reveals another layer of the difficulty in forming Uranus and Neptune. While formation models already showed that getting the required masses of composition of the planets within the disk lifetime requires special conditions (see \citealt{2020arXiv200710783H} and references therein) this study adds the importance of fitting other physical properties including their rotation rate, obliquity, and satellite system. 
The measured obliquities and rotation periods of Uranus and Neptune 
pose an additional challenge to planet  formation and evolution scenarios.

\section*{Acknowledgements}
This work has been carried out within the framework of the National Centre of Competence in Research PlanetS, supported by the Swiss National Foundation.
R.H.~acknowledges support from SNSF grant 200021\_169054. 
A.C.~acknowledges support from Forschungskredit of the University of Zurich, grant no. [FK-76104-01-01].
% We also thank the anonymous referee for valuable suggestions and comments that helped to improve the paper.
A.I.~acknowledges NASA grant 80NSSC18K0828 to Rajdeep Dasgupta.
The simulations were performed using the UZH HPC allocation on the Piz Daint supercomputer at the Swiss National Supercomputing Centre (CSCS). 

\section*{Data availability}
The data underlying this article will be shared on reasonable request to the corresponding author.

\begin{appendix}
\section{The effect of the mass ratio on the system's angular momentum}
\label{appendix:angularmomentum}
Here we briefly discuss why equal mass mergers of such massive embryos contain much angular momentum, and hence lead to short rotation periods. More generally for the same given colliding mass, an equal-mass collision contains more angular momentum than a higher target-to-impactor mass ratio. For simplicity we assume that the plane of collision coincided with the x-y plane.
Assuming no initial rotation the angular momentum of a collision is given by
\begin{equation}
    %L_z=b(R_t+R_i)v_{imp}(M_t\gamma^2+M_i(1-\gamma))^2,
    L_z=b \left( R_1 + R_2 \right) v_{imp} \left( M_1 \gamma^2 + M_2 \left(1-\gamma\right)^2 \right)
\end{equation}
where $R_1$, $R_2$ are the radii and $M_1$ and $M_2$ the masses of the two colliding bodies, $b$ is the impact parameter (or the sinus of the impact angle), $v_{imp}$ the impact velocity, and $\gamma=M_2/(M_1+M_2)$.
Using the total colliding mass $M=M_1+M_2$ and impact velocity this expression can be reduced to
\begin{equation}
    L_z=b \left( R_1 + R_2 \right) v_{imp} M \gamma \left(1-\gamma\right)
\end{equation}
where the bodies' radii
$R_1 = (3M/4\pi\rho_1)^{1/3} \gamma^{1/3}$
and
$R_2 = (3M/4\pi\rho_2)^{1/3} (1-\gamma)^{1/3}$
are only weakly depending on $\gamma$ and can be assumed to be constant.
For the same colliding mass and collision geometry, $L_z$ takes the highest value for $\gamma=1/2$, i.e., for equal-mass mergers. If the mass is distributed differently, for example in case of a 10:1 mass ratio (so $\gamma=1/11$), $L_z$ is about 30\% of the angular momentum contained in an equal mass merger involving the same total colliding mass. 

\section{Comparison to the N-body results}
\begin{figure}
	\includegraphics[keepaspectratio,width=0.45\textwidth]{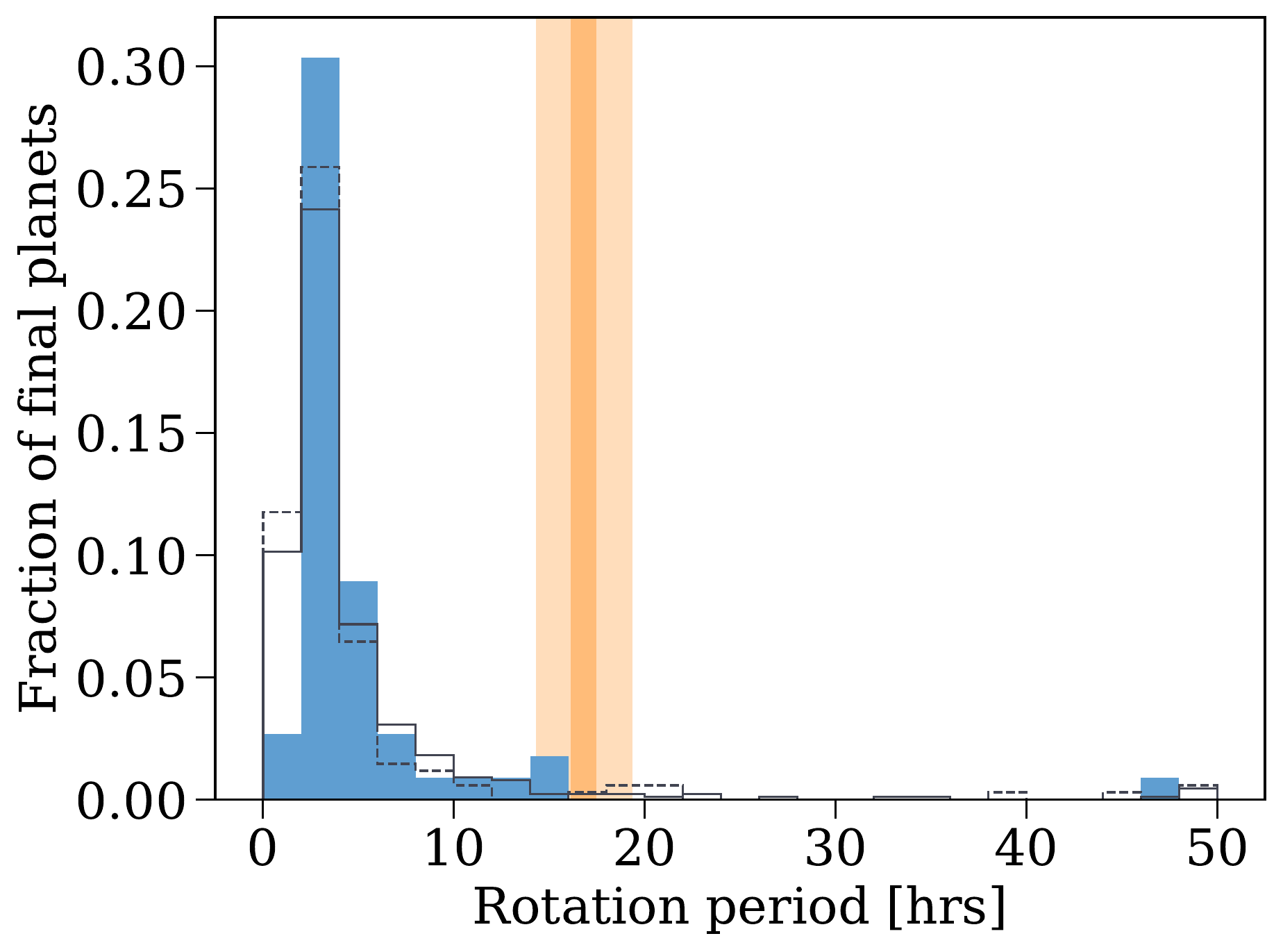}
    	\caption{   \textbf{The inferred distribution of the planet's final rotation periods.}
    The dark orange line represents values that are consistent with Uranus' and Neptune's inferred rotation periods and the light orange bar indicates values that agree with the mean value within 15\%. The dark grey lines show the inferred distribution from two different sets of I15.
    \label{fig:rotationperiod_hist_i15}
    	}
\end{figure}

Figure \ref{fig:rotationperiod_hist_i15} shows the distribution of the planet's inferred final rotation periods from our subset of simulations compared to two larger sets of simulations from I15. Our subset includes only systems with good U/N analogues while the I15 sets include all bodies that suffered at least one collision. The rotation periods for I15 are inferred assuming a two-body approximation, angular momentum conservation and a final spherical body of uniform density. The results are similar if the final body is modelled as a MacClaurin ellipsoid. The general trend for the rotation periods is similar in our SPH simulations and in I15, with $\sim$ 35\% of planets rotating very fast ($P\leq$ 5 hours), and only a few percent of the planets rotate with values close to the planets' current ones. The rotation periods obtained from the scenario of I15 are systematically higher than the ones inferred from our 3D hydro simulations with many of the planets   rotating very fast ($P\leq$ 2 hours). 

\section{Disk alignement}
\begin{figure*}
	\centering
	\includegraphics[keepaspectratio,width=0.95\textwidth]{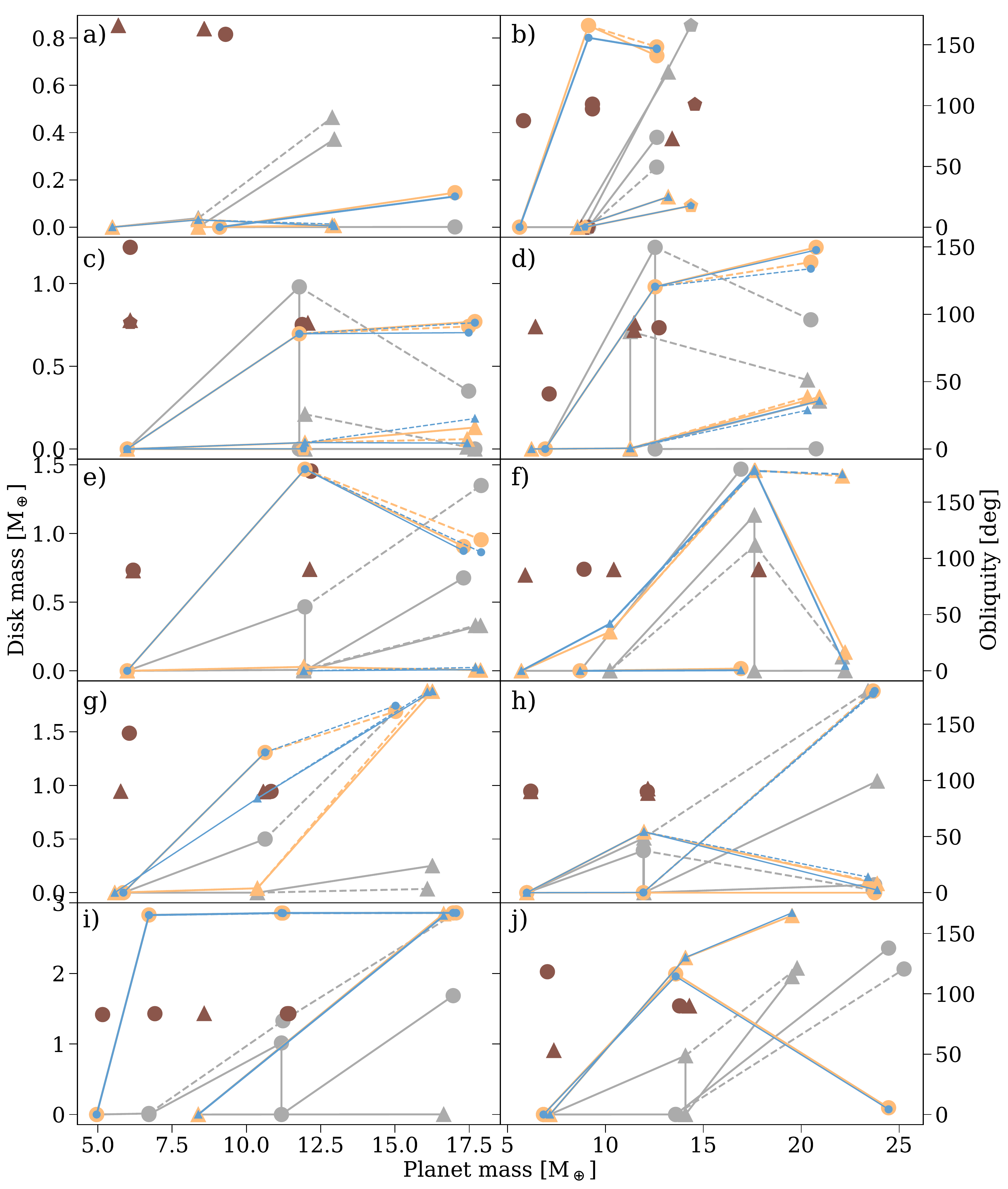}
    	\caption{\textbf{The planet-disk obliquity (blue and orange respectively) and the disk mass (gray) evolution}. Every panel represents a set of simulations. The symbols distinguish the two or three planets. The line follows the case when assuming a condensed planet as initial stated while the dashed line follows the case with a preexisting disk. In brown is over-plotted the obliquity of the initial impact velocity for that collision. \label{fig:disk_evolution_all}.
    	}
\end{figure*}

Figure \ref{fig:disk_evolution_all} shows the planet-disk obliquity and the disk mass after each collision for 10 sets of simulations where at least one of the planets formed a massive disk. 

%%%%%%%%%%%%%%%%%%%%%%%%%%%%%%%%%%%%%%%%%%%%%%%%
% Table of all simulations
%%%%%%%%%%%%%%%%%%%%%%%%%%%%%%%%%%%%%%%%%%%%%%%%
\clearpage
\onecolumn
\small{
\begin{longtable}{llllllllll}
\caption{\textbf{Table of all the simulations.} The capital letters list all the performed selected runs from I15. The second and third characters stand for the number of the planet and the number of the collision. The fifth character stands for the assumed initial thermal state of the planet c) condensed e) expanded. The asterisk indicates collisions that do not converge. The double asterisk indicates one second collision that was not performed as the first one did not converge. \label{table:allsims}} \\
\toprule
ID & $m_1$ [\ME]  & $m_2$ [\ME] & $v$ {[}km/s{]} & $m_{tot}$ [\ME] & $m_d$ [\ME] & $m_r/m_d$ & $P$ [hrs] & Obliquity [deg] & Disk obliquity [deg] \\ \toprule
A11c   & 5.99  & 5.99  & 17.41 & 11.97 & 0.466 & 0.014 & 3.11  & 178.8 & 178.9 \\
A12e   & 11.97 & 5.99  & 17.18 & 17.90 & 1.350 & 0.043 & 2.97  & 116.3 & 105.2 \\
A12c   & 11.97 & 5.99  & 17.18 & 17.88 & 1.210 & 0.076 & 3.05  & 113.4 & 106.4 \\
A21c   & 5.99  & 5.99  & 17.40 & 11.93 & 0.007 & 0.000 & 5.81  & 3.5   & 3.0   \\
A22e   & 11.93 & 5.99  & 15.88 & 17.71 & 0.330 & 0.000 & 3.33  & 0.7   & 1.1   \\
A22c   & 11.93 & 5.99  & 20.18 & 17.88 & 0.330 & 0.008 & 3.14  & 0.6   & 0.4   \\
\midrule
B11c   & 5.99  & 5.99  & 17.52 & 11.97 & 0.801 & 0.002 & 3.05  & 0.6   & 0.7   \\
B12c   & 5.99  & 5.99  & 17.42 & 11.94 & 0.392 & 0.000 & 2.76  & 0.0   & 0.2   \\
B13c   & 11.97 & 11.94 & 14.48 & 23.67 & 0.024 & 0.005 & 11.15 & 179.6 & 177.1 \\
B13e   & 11.97 & 11.94 & 21.96 & 23.75 & 0.074 & 0.002 & 5.69  & 0.0   & 179.6 \\
B21c   & 5.99  & 5.99  & 17.45 & 11.94 & 0.006 & 0.000 & 4.60  & 180.0 & 177.3 \\
B22c   & 5.99  & 5.99  & 17.75 & 11.97 & 0.510 & 0.018 & 3.17  & 54.1  & 54.2  \\
B23e   & 11.94 & 11.97 & 14.49 & 23.41 & 1.880 & 0.028 & 2.93  & 10.8  & 13.9  \\
B33c  & 11.94 & 11.97 & 21.97 & 23.88 & 1.040 & 0.000 & 2.46  & 8.1   & 2.3   \\
\midrule
C11c   & 5.99  & 5.99  & 17.47 & 11.78 & 0.980 & 0.000 & 2.98  & 82.1  & 82.1  \\
C12e   & 11.78 & 5.99  & 16.15 & 17.48 & 0.350 & 0.002 & 5.22  & 87.3  & 82.9  \\
C12c   & 11.78 & 5.99  & 20.16 & 17.69 & 0.000 & 0.000 & 6.08  & 90.8  & 90.0  \\
C22c   & 5.99  & 5.99  & 17.49 & 11.97 & 0.210 & 0.000 & 3.19  & 4.5   & 4.6   \\
C22e   & 11.97 & 5.99  & 14.84 & 17.43 & 0.012 & 0.000 & 14.22 & 7.0   & 4.2   \\
C22c   & 11.97 & 5.99  & 20.17 & 17.69 & 0.000 & 0.000 & 46.24 & 15.3  & 21.7  \\
C31c   & 5.99  & 5.99  & 17.74 & 11.92 & 1.260 & 0.000 & 2.85  & 0.0   & 0.1   \\
\midrule
D11c   & 5.86  & 4.78  & 16.83 & 10.63 & 0.500 & 0.004 & 2.99  & 124.9 & 125.0 \\
D12e   & 10.63 & 4.91  & 14.29 & 15.02 & 1.692 & 0.047 & 2.83  & 161.5 & 166.6 \\
D12c*  & 10.63 & 4.91  & 19.26 & -     & -     & -     & -     & -     & -     \\
D21c   & 5.57  & 4.85  & 16.73 & 10.37 & 0.000 & 0.000 & 24.12 & 3.9   & 84.0  \\
D22e   & 10.37 & 5.96  & 14.54 & 16.06 & 0.035 & 0.002 & 3.77  & 179.7 & 178.8 \\
D22c   & 10.37 & 5.97  & 19.45 & 0.50  & 0.251 & 0.006 & 3.37  & 179.4 & 179.5 \\
\midrule
E11c   & 5.62  & 3.56  & 16.08 & 9.14  & 0.000 & 0.000 & 9.61  & 165.7 & 155.8 \\
E12e   & 9.13  & 3.63  & 14.43 & 12.62 & 0.254 & 0.004 & 3.46  & 148.3 & 147.0 \\
E12c   & 9.13  & 3.63  & 19.36 & 12.63 & 0.380 & 0.074 & 3.49  & 147.9 & 146.5 \\
E21c   & 8.58  & 4.65  & 18.10 & 13.21 & 0.656 & 0.024 & 2.99  & 24.9  & 24.9  \\
E31c   & 8.96  & 5.71  & 18.69 & 14.37 & 0.853 & 0.068 & 2.94  & 17.6  & 17.7  \\
\midrule
F11c   & 9.70  & 9.34  & 20.38 & 18.74 & 2.520 & 0.058 & 2.69  & 61.6  & 61.7  \\
F21c   & 8.71  & 7.78  & 19.43 & 16.44 & 1.298 & 0.058 & 2.48  & 160.2 & 160.2 \\
F22e*  & 16.49 & 2.68  & 17.24 & -     & -     & -     & -     & -     & -     \\
F22c*  & 16.49 & 2.68  & 20.53 & -     & -     & -     & -     & -     & -     \\
\midrule
G11c   & 9.48  & 9.10  & 20.16 & 18.57 & 1.413 & 0.002 & 2.63  & 30.9  & 30.9  \\
G21c   & 8.02  & 6.30  & 16.95 & 14.19 & 1.434 & 0.068 & 3.23  & 0.2   & 0.3   \\
G22e   & 14.32 & 2.43  & 7.87  & 16.21 & 1.922 & 0.089 & 4.04  & 21.7  & 20.2  \\
%G22c & 14.32 & 2.43  & 18.22 &       &       &       &       &       &       \\
\midrule
H11c   & 7.65  & 1.30  & 16.43 & 8.63  & 0.013 & 0.192 & 5.49  & 169.8 & 170.5 \\
H12e   & 8.95  & 7.99  & 15.72 & 16.58 & 2.077 & 0.150 & 2.58  & 46.9  & 46.3  \\
H12c   & 8.95  & 7.99  & 16.03 & 16.03 & 1.600 & 0.147 & 3.11  & 46.2  & 45.9  \\
H21c*  & 7.37  & 4.64  & 17.53 & -     & -     & -     & -     & -     & -     \\
H22**  & 12.01 & 4.30  & -     & -     & -     & -     & -     & -     & -     \\
\midrule
I11c   & 9.58  & 8.00  & 19.77 & 17.56 & 0.596 & 0.026 & 2.80  & 0.1   & 0.1   \\
I12e   & 17.58 & 11.07 & 14.12 & 28.24 & 0.029 & 0.026 & 4.00  & 179.9 & 178.6 \\
I12c   & 17.58 & 11.07 & 23.56 & 28.53 & 0.342 & 0.006 & 3.15  & 180.0 & 179.8 \\
I21c   & 8.02  & 7.87  & 19.24 & 15.87 & 0.434 & 0.005 & 2.88  & 179.9 & 179.7 \\
I22e   & 15.89 & 7.86  & 14.53 & 23.29 & 1.460 & 0.121 & 7.98  & 0.2   & 0.1   \\
I22c*  & 15.89 & 7.86  & 22.11 & -     & -     & -     & -     & -     & -     \\
\midrule
J11c   & 6.84  & 6.78  & 18.16 & 13.59 & 0.003 & 0.003 & 3.88  & 116.5 & 114.5 \\
J12e   & 13.62 & 11.71 & 17.78 & 25.25 & 2.063 & 0.000 & 2.43  & 7.5   & 8.2   \\
J12c   & 13.62 & 11.71 & 22.47 & 24.46 & 2.359 & 0.202 & 2.95  & 5.6   & 4.3   \\
J21c   & 7.17  & 7.02  & 18.46 & 14.09 & 0.834 & 0.017 & 2.61  & 130.0 & 130.1 \\
J22e   & 14.19 & 5.79  & 14.13 & 19.79 & 2.075 & 0.018 & 2.63  & 163.8 & 156.3 \\
J22c   & 14.19 & 5.79  & 20.73 & 19.53 & 1.955 & 0.123 & 3.55  & 164.8 & 167.1 \\
\midrule
K11c   & 4.96  & 1.82  & 14.58 & 6.72  & 0.014 & 0.004 & 5.56  & 178.1 & 178.0 \\
K12e   & 6.78  & 4.56  & 13.20 & 11.23 & 1.329 & 0.055 & 3.06  & 179.8 & 179.7 \\
K12c   & 6.78  & 4.56  & 16.72 & 11.18 & 1.014 & 0.048 & 2.92  & 179.8 & 179.9 \\
K13e   & 11.34 & 6.87  & 7.73  & 17.06 & 2.859 & 0.196 & 2.84  & 180.0 & 179.9 \\
K13c   & 11.34 & 6.87  & 18.25 & 16.96 & 1.685 & 0.104 & 4.73  & 179.9 & 180.0 \\
K21c   & 8.38  & 8.31  & 19.43 & 16.64 & 0.002 & 0.000 & 4.21  & 179.1 & 177.4 \\
\midrule
L11c   & 9.10  & 7.97  & 19.56 & 17.02 & 0.001 & 0.000 & 7.18  & 15.6  & 13.9  \\
L21c   & 5.49  & 2.89  & 15.55 & 8.38  & 0.038 & 0.000 & 4.64  & 3.4   & 3.3   \\
L22e   & 8.38  & 4.63  & 13.66 & 12.89 & 0.464 & 0.008 & 3.18  & 0.8   & 1.4   \\
L22c   & 8.38  & 4.63  & 18.02 & 12.96 & 0.372 & 0.007 & 3.06  & 0.7   & 0.5   \\
\midrule
M11c   & 6.93  & 5.95  & 17.89 & 12.54 & 1.219 & 0.039 & 2.72  & 120.4 & 120.5 \\
M12e   & 12.89 & 8.30  & 15.15 & 20.49 & 0.781 & 0.034 & 5.48  & 138.7 & 133.8 \\
M12c   & 12.89 & 8.30  & 21.22 & 20.76 & 0.001 & 0.000 & 5.18  & 149.7 & 147.7 \\
M21c   & 6.23  & 5.06  & 17.52 & 11.27 & 0.710 & 0.003 & 2.86  & 0.5   & 0.6   \\
M22e   & 11.29 & 9.73  & 14.11 & 20.32 & 0.417 & 0.056 & 3.25  & 38.6  & 28.8  \\
M22c   & 11.29 & 9.73  & 21.01 & 20.93 & 0.290 & 0.000 & 2.97  & 38.7  & 35.8  \\
\midrule
N11c   & 8.30  & 5.85  & 18.45 & 14.16 & 0.673 & 0.085 & 3.11  & 81.5  & 81.4  \\
N21c   & 8.96  & 3.57  & 17.96 & 12.42 & 0.001 & 0.000 & 8.02  & 177.6 & 168.4 \\
\midrule
O11c   & 8.31  & 6.78  & 18.85 & 14.63 & 1.529 & 0.194 & 3.64  & 113.8 & 113.8 \\
O21c   & 8.02  & 7.14  & 18.79 & 15.17 & 0.135 & 0.000 & 3.40  & 70.7  & 70.9  \\
\midrule
P11c   & 6.54  & 5.90  & 17.60 & 12.35 & 0.002 & 0.000 & 6.23  & 154.2 & 154.4 \\
%P21c & 9.99  & 4.65  &       &       &       &       &       &       &       \\
\midrule
Q11c   & 9.90  & 9.29  & 18.85 & 19.23 & 0.662 & 0.014 & 2.73  & 3.3   & 3.3   \\
Q21c   & 8.52  & 6.84  & 18.92 & 15.42 & 0.033 & 0.000 & 3.92  & 83.4  & 83.2  \\
\midrule
R11c   & 8.71  & 8.25  & 19.63 & 16.92 & 1.469 & 0.040 & 2.44  & 2.0   & 0.5   \\
R21c   & 5.71  & 4.57  & 16.65 & 10.23 & 0.000 & 0.000 & 12.71 & 34.7  & 41.9  \\
R22e   & 10.28 & 7.46  & 15.40 & 17.65 & 0.915 & 0.024 & 2.91  & 178.2 & 177.7 \\
R22c   & 10.78 & 7.46  & 19.03 & 17.61 & 1.134 & 0.041 & 2.67  & 178.2 & 179.3 \\
R32e   & 17.74 & 5.03  & 8.63  & 22.10 & 0.102 & 0.005 & 15.62 & 173.2 & 174.8 \\
R32c   & 17.74 & 5.03  & 21.98 & 22.24 & 0.001 & 0.000 & 12.52 & 16.5  & 4.4   \\
\midrule
S11c   & 7.20  & 6.99  & 15.55 & 14.15 & 0.006 & 0.000 & 4.75  & 4.4   & 1.2   \\
S12c   & 7.37  & 7.00  & 18.29 & 14.24 & 1.830 & 0.052 & 2.70  & 37.8  & 37.9  \\
%S22c & 14.37 & 5.57  &       &       &       &       &       &       &
\toprule
\end{longtable}
}
\clearpage
\twocolumn

\end{appendix}

\clearpage
\bibliography{bib_uranus_neptune}{}
\bibliographystyle{mnras}
\end{document}